# Laser Fault Injection Attacks against Radiation Tolerant TMR Registers

Dmytro Petryk[1], Zoya Dyka[1], Ievgen Kabin[1], Anselm Breitenreiter[1], Jan Schäffner[1] and Milos Krstic[1,2]

[1] *IHP – Leibniz-Institut für innovative Mikroelektronik,* Frankfurt (Oder), Germany
[2] *University of Potsdam,* Potsdam, Germany
{petryk, dyka, kabin, breitenreiter, schaeffner, krstic}@ihp-microelectronics.com

Security requirements for the Internet of things (IoT), wireless sensor nodes, and other wireless devices connected in a network for data exchange are high. Confidentiality, data integrity, availability of the services, and authentication of communicated devices are the main security goals usually implemented with cryptographic approaches. Their secrecy is based on the used keys. Theoretically, the cryptographic approaches are secure. This changes dramatically if an attacker has physical access to the attacked device. The execution time of cryptographic operations, current drawn from the power supply, electromagnetic radiation and many other physical effects, called side-channel effects, can be measured and later analysed using statistical or machine learning methods. Many attacks concentrate on the analysis of registers' activity. Some attacks exploit the key-dependent hamming weight of the data, stored in registers, and require usually many traces for the analysis [1]. Other attacks exploit the distinguishability of the registers addressing and can be successful when analysing many traces [2] as well as a single trace only [3], [4]. The sensitivity of cryptographic chips to the environmental and operating parameters – temperature, voltage, frequency, light, EM pulses, radiation, etc. – can be successfully exploited to reveal the cryptographic key too. A fluctuation of the environment and working parameters can cause fault(s). Cryptographic keys can be revealed by analysing such faults, whereby the mostly attacked blocks of cryptographic designs are the registers. Optical FI attacks using lasers belong to the class of semi-invasive attacks. This type requires to perform a chip decapsulation and exploits the sensitivity of semiconductors to the visible light. The pioneering work regarding optical FI attacks was published already in 2002 [5].

During our previous works we successfully attacked different standard logic gates manufactured in IHP's technologies [6], [7]. Additionally, we investigated the resistance of Resistive Random Access Memory (RRAM) cells [13] and radiation hard Junction Isolated Common Gate (JICG) FF [8]. A short overview of the optical FI attacks and used equipment is given in [9]. A common approach to increase resistance against FI attacks is to implement countermeasures based on redundancy techniques. In this work, we investigate the robustness of the Triple Modular Redundancy (TMR) shift registers against optical FI attacks on example of the TMR shift registers based on standard TMR architecture, designed and manufactured in the IHP's CMOS 130 nm technology [14]. TMR flip-flops attacked have been originally designed for space application, have been evaluated in single event effect measurements, and have been verified as very radiation tolerant. Each attacked TMR circuit is a 1024-bit long register, i.e. it contains 1024·3= 3072 standard library flip-flops and 1024 voters. The block diagram of the attacked TMR-FF is shown in **Fig. 1**.

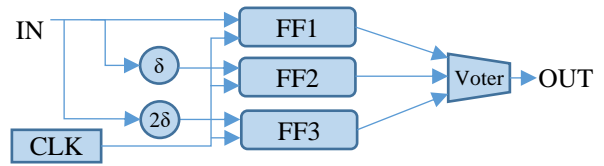

**Fig. 1.** Block diagram of attacked TMR-FF.

The IHP TMR-FF has two delay elements denoted in **Fig. 1** as $\delta$ and $2\delta$. These delay elements have to filter possible transient faults caused by a particle at the inputs of the FFs. Detailed discussion of the chosen delay value can be found in [14]. According to the block diagram of TMR-FF, a FI attack can be successful if a fault will be injected into any two FFs of the same TMR-FF simultaneously, or if a fault will be injected into logic gates of the voter (and it is longer then the timing filter limits). Due to the technology requirements the chips have metal fillers atop.

To attack the IHP radiation hard TMR shift registers we used a setup that consists of: a modified Riscure Diode Laser Station (DLS) [11], a VC glitcher, a PC with the Riscure Inspector software, a stable power supply, a generator and an oscilloscope. The setup for optical FI attacks is shown in **Fig. 2** schematically.

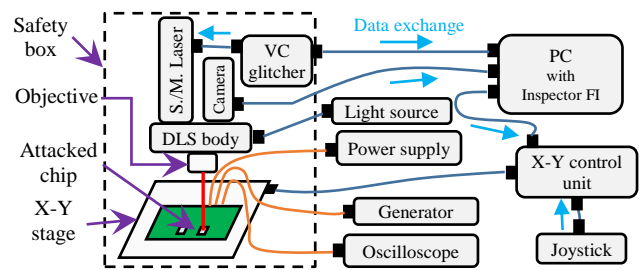

**Fig. 2.** Optical Fault Injection setup.

The modified FI setup can be used with two multi-mode lasers: red (808 nm) and near-infrared (1064 nm) laser [11] and with a single-mode red (808 nm) laser from Alphanov [12]. Detailed description of the setup, including the applied laser parameters, can be found in [6]–[8]. Due to the fact that the attacked chip was manufactured by the IHP, we did not perform a decapsulation and we decided to attack the register from the front-side of the chips.

A triplicated flip-flop and its voter in attacked registers are placed in one row. Details about placement of gates and distances between them can be found in [10] and [14]. Due to the specific placement of the gates, we have three possible attack scenarios:

*Scenario1*: Attack illuminating logic gates of the voter: successful injection of a fault into any logic gate of the voter can lead to a faulty logic state at voters' output that is the output of the TMR-FF. This scenario is technically possible using both red lasers available in the setup.

*Scenario2*: Attack illuminating FF1 and FF2 simultaneously: this scenario is possible using the multi-mode laser with 20× or 5× magnification objectives only.

*Scenario3*: Attack illuminating the whole TMR-FF, i.e. illuminating 3 FFs and the voters simultaneously: the scenario is technically possible using the multi-mode laser with 5× magnification objective only. We did not verify this scenario due to the possible simultaneous influence on FFs and voter that leads to a fault-free state.

To observe a successful fault injection we sent a constant input to the attacked shift register during an experiment, i.e. the input of the shift register was set to '1' or to '0'. We obtained the following results:

- In *Scenario1* all our FI attacks illuminating the voter of different FFs were not successful despite successful injection of faults into standard library FFs [7]. This result can be explained due to the difference regarding laser parameters between experiments described in [7] and here: in previous work we used a standard DLS from Riscure, after then the DLS was modified by Opto GmbH to allow using the Alphanov single-mode laser. We assume that the modification reduced the laser beam power.
- In *Scenario2* we were able to inject faults using the multi-mode laser with a 20× and 5× magnification objectives. We observed transient repeatable faults in flip-flops in logical state '0' and logical state '1', i.e. bit-set ('0'→'1') and bit-reset ('1'→'0') faults. We assume that we influenced on both FFs simultaneously. Please note that not all TMR-FFs attacked where successfully influenced. This can be due to the metal fillers placed over the sensitive areas of TMR-FFs. In our experiments, we did not observe any permanent (change of logic state is no more possible) or stuck-at (change of logic state cannot be changed before the device shutdown) faults, even when applying maximum laser beam power and pulse duration. After the successful injection of faults, the TMR shift register is fully functional, i.e. the internal structure of the chip is intact and its surface was not damaged. TABLE I gives a quantitative overview of our attacks for *Scenario2*.

Generally, redundancy is a means to increase the resistance of a design to faults and may be a promising approach against registers manipulations by cryptographic devices. But in our experiments, we were able to inject different transient faults into TMR registers using a single laser beam. Performing attacks against FFs we were able to inject transient repeatable *bit-set* as well as *bit-reset* faults into the IHP TMR shift registers using the multi-mode laser.

TABLE I. RESULTS OF ATTACKS AGAINST TMR-REGISTERS USING MULTI-MODE LASER FOR SCENARIO2

| Register clock frequency, MHz | Register input | Power, %[a] | Pulse, ns | Magnification objective | Type of fault |
|---|---|---|---|---|---|
| 10 | '0' '1' | 40-100 45-100 | 130-280 | 20× | bit-set bit-reset |
| 10 | '0' '1' | 65-100 55-100 | 130-280 | 5× | bit-set bit-reset |
| 50 | '0' '1' | 80-100 90-100 | 80 | 20× | bit-set[b] bit-reset |
| 50 | '0' '1' | 90-100 60-100 | 50-80 | 5× | bit-set bit-reset |

[a.] POWER MEASUREMENT UNIT IN RISCURE SOFTWARE.
[b.] THE FAULT IS NOT FULLY REPEATABLE.

REFERENCES

[1] P. Kocher, J. Jaffe, and B. Jun, "Differential Power Analysis", in Advances in Cryptology — CRYPTO' 99, vol. 1666, Berlin, Heidelberg: Springer Berlin Heidelberg, 1999, pp. 388–397.

[2] K. Itoh, T. Izu, and M. Takenaka, "Address-Bit Differential Power Analysis of Cryptographic Schemes OK-ECDH and OK-ECDSA", in Cryptographic Hardware and Embedded Systems - CHES 2002, Aug. 2002, pp. 129–143.

[3] I. Kabin, Z. Dyka, D. Klann, and P. Langendoerfer, "Methods increasing inherent resistance of ECC designs against horizontal attacks", Integration, vol. 73, Jul. 2020, pp. 50–67.

[4] I. Kabin, Z. Dyka, and P. Langendoerfer, "Atomicity and Regularity Principles Do Not Ensure Full Resistance of ECC Designs against Single-Trace Attacks", Sensors, vol. 22, no. 8, Art. no. 8, Jan. 2022.

[5] S. Skorobogatov and R. Anderson, "Optical Fault Induction Attacks", Workshop on Cryptographic Hardware and Embedded Systems (CHES), USA, San Francisco, Aug, 13–15, 2002, pp. 2–12.

[6] D. Petryk, Z. Dyka and P. Langendörfer, "Sensitivity of Standard Library Cells to Optical Fault Injection Attacks in IHP 250 nm Technology", 2020 9th Mediterranean Conference on Embedded Computing (MECO), Montenegro, Budva, June 8–11, 2020, pp. 1–4.

[7] D. Petryk, Z. Dyka, J. Katzer and P. Langendörfer, "Metal Fillers as Potential Low Cost Countermeasure against Optical Fault Injection Attacks", 2020 IEEE East-West Design & Test Symposium (EWDTS), Bulgaria, Varna, Sept. 4–7, 2020, pp. 1–6.

[8] D. Petryk, Z. Dyka, R. Sorge, J. Schäffner and P. Langendörfer, "Optical Fault Injection Attacks against Radiation-Hard Shift Registers", 2021 24th Euromicro Conference on Digital System Design (DSD), Italy, Palermo, Sept. 1–3, 2021, pp. 371–375.

[9] D. Petryk, Z. Dyka, P. Langendörfer, "Optical Fault Injections: a Setup Comparison", Proc. PhD Forum of the 8th BELAS Summer School, Estonia, Tallinn, June 20–22, 2018, pp. 1–5.

[10] V. Petrovic and M. Krstic, "Design Flow for Radhard TMR Flip-Flops", 2015 IEEE 18th International Symposium on Design and Diagnostics of Electronic Circuits & Systems (DDECS), Serbia, Belgrade, Apr. 22–24, 2015, pp. 203–208.

[11] Riscure. Diode Laser Station Datasheet, 2011. https://www.riscure.com/security-tools/inspector-hardware/

[12] Alphanov PDM laser sources. Rise time comparison between a PDM HPP at 1064 and 980nm and a standard PDM+. URL: https://www.alphanov.com/en/products-services/pdm-laser-sources

[13] D. Petryk, Z. Dyka, E. Perez, I. Kabin, J. Katzer, J. Schäffner and P. Langendörfer, "Sensitivity of HfO$_2$-based RRAM Cells to Laser Irradiation", Microprocessors and Microsystems, Volume 87, 2021, 104376, ISSN 0141-9331, pp. 1–20.

[14] O. Schrape, M. Andjelković, A. Breitenreiter, S. Zeidler, A. Balashov and M. Krstić, "Design and Evaluation of Radiation-Hardened Standard Cell Flip-Flops", in IEEE Transactions on Circuits and Systems I: Regular Papers, vol. 68, no. 11, Nov. 2021, pp. 4796-4809.